\documentclass[aps,twocolumn,showpacs,preprintnumbers,amsmath,amssymb,floatfix]{revtex4}
\usepackage{graphicx, epsfig}
\usepackage{bm}
\usepackage{amssymb}
\usepackage{amsmath}
\usepackage{amsfonts}
\newcommand{\be}{\begin{equation}}
\newcommand{\ee}{\end{equation}}
\newcommand{\bea}{\begin{eqnarray}}
\newcommand{\eea}{\end{eqnarray}}

\newcommand{\kt}{\rangle}
\newcommand{\br}{\langle}

\newcommand{\ed}{\end{document}}

\begin{document}

\title{Noise-induced   Synchronization in  Small World Networks of Phase Oscillators }

\author{ Reihaneh Kouhi Esfahani}
\author{ Farhad Shahbazi} 
\author{Keivan  Aghababaei Samani} 
\affiliation{ Department of Physics, Isfahan University of
Technology, Isfahan 84156-83111, Iran}

\begin{abstract}
A small-world network (SW) of similar phase oscillators, interacting according to the Kuramoto model
is studied numerically.  It is shown that deterministic Kuramoto dynamics on the SW networks has
various  stable stationary  states. This  can be attributed to the so called {\it  defect patterns } in a SW network which is inherited to it from deformation of {\it helical  patterns} in its  regular parent. Turning on an uncorrelated random force, causes the vanishing of the defect patterns, hence
increasing  the synchronization among oscillators for moderate noise intensities. This phenomenon called {\it stochastic synchronization} is generally  observed
in some natural networks like brain neural  network.
\end{abstract}

\maketitle

\section{Introduction}

 Noise is usually considered as a source of disturbance against the
main signals in laboratory as well as natural systems.
Nevertheless, the interplay between the randomness, created  by
the noise, and non-linearities may lead to the enhancement of
regular behavior in some dynamical systems~\cite{order-noise}.
Stochastic resonance~\cite{SR}, coherence resonance~\cite{CR},
noise-induced transport~\cite{transport}, noise-induced
transition~\cite{transition} and noise-induced collective firing
in excitable media~\cite{collective-firing} are  examples of such
a novel phenomena. Being noisy, nature takes the advantages of
these mechanisms  in order to   employ the random fluctuation as
agent of self-organization. This is the main reason that why
living systems work so reliable in spite of the presence of
various sources of  noise.

Brain neurons are examples of  biological systems  in which the
source of random fluctuations is the background synaptic noises
caused by highly fluctuating inputs coming from thousands of
other neurons connected to a given neuron~\cite{noise-neuron}.
However, this noise plays a constructive role in regular spiking
of the individual neurons and also increasing the synchronization
among the clusters of connecting neurons~\cite{brain-sync}.
Synchronous spiking  among a subset of neurons plays an important
role in more efficient propagation of  activities from a group of
neurons to another~\cite{signal-neuron}. Furthermore, there are
also some  controversial idea on  encoding of information about
stimuli thorough synchrony in oscillatory activity of
neurons~\cite{code-sync}.  Another phenomenon  in which
noise-induced synchronization takes place, is gene regulatory
process in systems such as quorum-sensing bacteria, in which noise
originates from the small number of molecules involved in the
related  biochemical reactions~\cite{noise-gene}.

Collective dynamical behaviors, like synchronization,  can be
observed  in  systems of coupled non-linear oscillators  and
extensively been studied on complex networks~\cite{rev-network}.
One of such models has been proposed by Kuramoto, which  consists
of a set of oscillators with fixed amplitude (phase oscillators)
mutually  coupled  by a $2\pi$ periodic interaction~\cite{kuramoto}. The stochastic Kuramoto model has been studied on the globally connected~\cite{skuramoto} and also on
scale-free (SF) and Erd\"{o}s-R\`{e}nyi (ER) random
networks~\cite{khoshbakht}. Analytical results on an all-to-all
network show that for a given distribution of intrinsic
frequencies of oscillators, a minimum value of coupling is needed
for synchronization. Perturbing  the fully synchronized state by an
uncorrelated white noise, causes the synchrony between oscillators
falls monotonically by increasing the noise strength. The same
results have been found on numerical integrations  of the
stochastic Kuramoto model on the ER and SF networks. The
difference is that the synchronized state in SF networks persists
more against applying the noise with respect to the ER and
all-to-all networks\cite{khoshbakht}.

Watts and Strogatz found out that many systems in nature
possess the properties of the small-world (SW)
networks~\cite{watts,strogatz}. Short mean path between the nodes
and high degree of  clustering are the two main features of SW
networks.  Former is a characteristic of random, while the latter
is  feature  of regular networks.  It has been found  that the
presence of random short-cuts,  may lead to noise driven ordering
phenomena such as stochastic resonance~\cite{sw-sr} and coherence
resonance~\cite{sw-cr} in SW networks.

Motivated by recent discoveries revealing  SW topology  of brain
neural networks~\cite{sw-brain} and also noise induced regulatory
behaviors in such a networks~\cite{brain-sync}, we study  the
effect of random force on the dynamics of SW network of a set of
similar  phase oscillators coupled   to each other based on
Kuramoto model. We will show that in this system, for intermediate
noise strength, the synchronization among the oscillators is
increased. The rest of the paper is organized as follows. First,
we present the results of numerical integration  of deterministic
Kuromato model on regular and SW networks.  Investigation of
Stochastic Kuramoto model driven by uncorrelated   white noise is
done in next section   and final section  is devoted to summary
and concluding remarks.

\section{Kuramoto model on complex networks}
In this section we introduce the Kuramoto model and numerically
investigate its steady state solutions on ER, SF and SW networks.
Consider a set of  phase oscillators, residing on  the top of the
nodes of a  network. Their phases and intrinsic oscillation
frequencies are given by $\theta_{i}$ and $\omega_{i}$,
respectively. According to the Kuramoto model, dynamics of these
phase oscillators is given by the following  set of coupled
differential equations:
  \be
    \dot{\theta}_i=\omega_i+K\sum_{j=1}^{N}a_{ij}
    \sin(\theta_j-\theta_i)\; ,i=1,\ldots,N\;,
    \label{kuramoto}
  \ee
where $K$ is the coupling strength, $ N $  is the number of nodes and $a_{ij}$ is the element of
adjacency matrix ($a_{ij}=1$ if nodes $i$ and $j$ are connected
and  $a_{ij}=0$ otherwise).

Synchronization of the Kuramoto model on SW networks, for random
distribution of $ \omega_{i} $ has already been studied by Hong.
et al~\cite{sw-synch}. They showed that small fraction of
shortcuts is enough for both phase and frequency synchronizations,
in spite of  absence of any synchronization on regular ones. In
the present work, we assume that all the intrinsic frequencies are
the same $(\omega_i=\omega_0)$, therefore moving   to a reference
frame in which $\omega_0=0$, simplifies Eq.(\ref{kuramoto}) to:
    \be
    \dot{\theta}_i=K\sum_{j=1}^{N}a_{ij}
    \sin(\theta_j-\theta_i)\;,i=1,\ldots,N.
    \label{kuramoto2}
    \ee

To compare the solutions of Kuramoto model in these three  networks, we need to construct
them with equal number of nodes and edges. For building  a  SF network with average
 connectivity $\langle k \rangle=2m$, we use the Barab\'{a}si-Albert  (BA) algorithm \cite{BA}.
Starting from $m_{0}$ initial connected nodes, one attaches a newly entering  node to
$m\leq m_{0}$ elder ones with  probability proportional to the degree of the present nodes.
An ER random network  with  $N$ nodes and  the same
average degree per node~($\langle k \rangle=2m$),  is simply produced by
connecting   randomly chosen pair of nodes with $Nm$ edges~\cite{ER}.
To construct the SW network, we use Watts-Strogatz (WS) algorithm~\cite{watts}.
Starting from a regular network with $N$ nodes  and $k=2m$ edges for each node,
we rewire each edge randomly with probability $p$. Choosing $0.005  \lesssim p\lesssim 0.05$,
this  process converts the initial regular network  to a complex network with a small mean path
length  and large  clustering coefficient, characteristics of SW networks.

Starting from a randomly distributed initial phases $\theta_{i}(0)$ (which is selected from a box distribution
in the interval $[-\pi,\pi[$~), the set of coupled differential Eqs.(\ref{kuramoto2}) are integrated from $t=0$
to a given time $t$ with the time step $dt$, using Euler method. This method enables us to compute $\theta_{i}(t)$ and to determine
the synchrony among the oscillators at any time, we define the following complex order parameter:
 \be re^{i\psi}=\frac{1}{N}\sum_{j=1}^{N}e^{i\theta_j(t)},
    \label{r}
 \ee
where $0\leq r(t)  \leq 1$ indicates the degree of synchronization in the network and $\psi$ is the  phase of the order parameter.
\begin{figure}
\begin{center}
\includegraphics[scale=0.4]{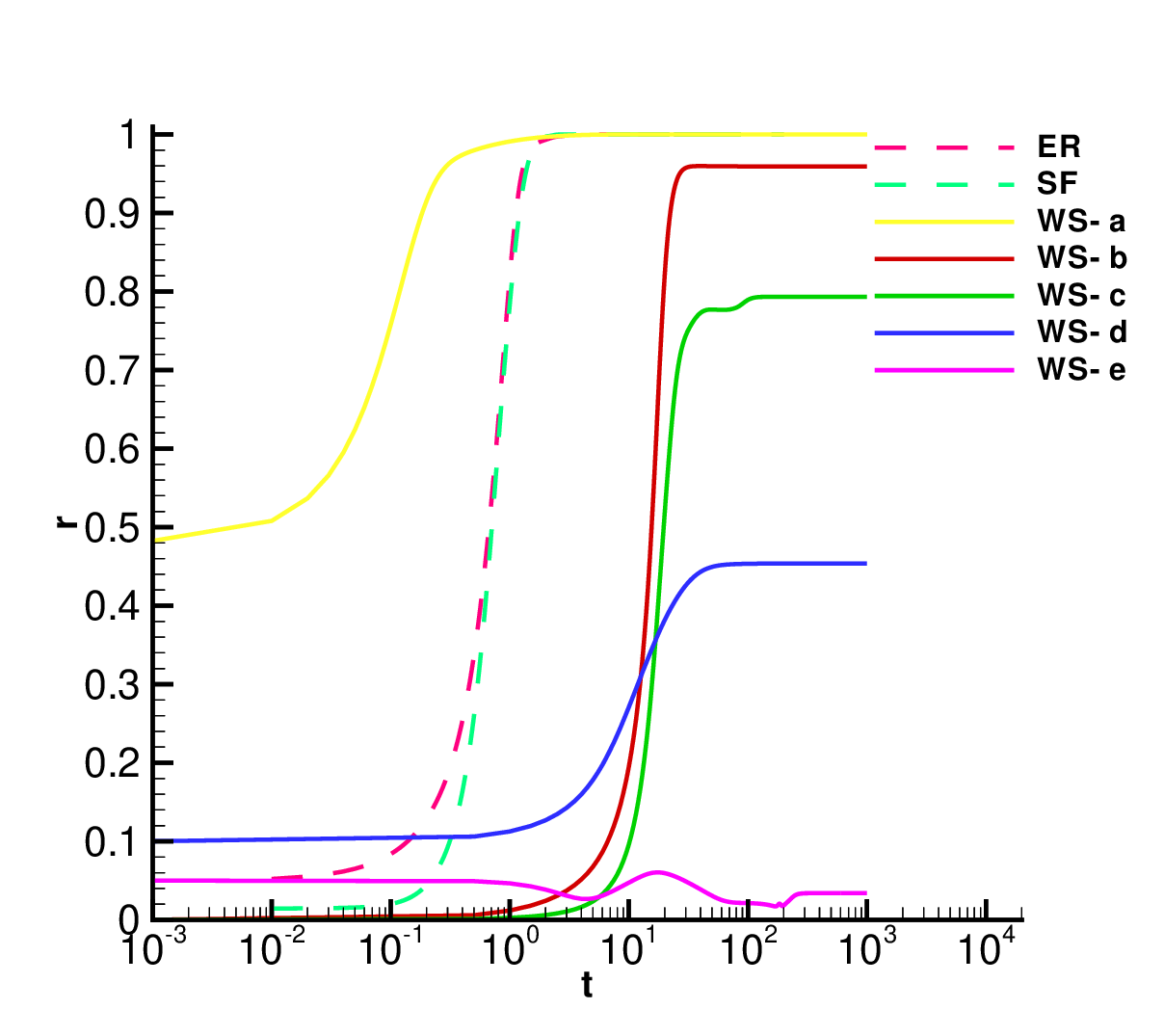}
\caption{(Color on-line) Order parameter ($r$) versus time (in
logarithmic scale) for ER (dashed dark-gray curve), BA (dashed light-gray curve) and WS  networks for five different initial conditions (solid curves from top to bottom for WS-a to WS-e, respectively). $N=1000$ and
$\langle k \rangle=10$.}
\label{fig1}
\end{center}
\end{figure}
Fig.(\ref{fig1}) shows the temporal variations of the $r(t)$  on
the three types of networks with $N=1000$ and $\langle k
\rangle=10$. To obtain these plots, time step is set to $dt=0.01$
and  five realizations of initial phase distributions are taken
for a fixed  network of each type. The rewiring probability for
constructing the WS network out of regular one is chosen to be $
p=0.04 $. As can be seen, the oscillators  on ER and BA networks
immediately reach  to a fully synchronized state ($r=1$)
irrespective to the initial conditions. However,  in the case of
WS network, they  more slowly go toward the steady states which
are highly dependent on the initial phase distributions in such a
way the $r(\infty)$  reaches several values between $0$ and $1$.
These results show that in contrast to ER and SF networks, the
structure of steady states of the Kuramoto model on  SW networks
can be  more complex. In what follows, we discuss that the
sensibility of dynamics  to initial conditions  is indeed
inherited to SW networks  from their  regular network parents.  In
a regular network,  the ratio of  nearest neighbor connections to
network size  ($ k/N $)  determines the number of stable
solutions. It has been shown that for $ k/N < 0.34 $,  different
initial conditions lead to different final
states~\cite{reg-basin}.

It is easy to show that the stable stationary solutions of Eqs.(\ref{kuramoto2})  have to satisfy the following conditions:
\be
\sum_{i=1}^{N}\sin\theta_{i}=\sum_{i=1}^{N}\cos\theta_{i}=0,
\ee
provided that the phase difference between any two adjacent oscillators be less than
$ \pi/2 $  ({\it i.e}, $ \Delta\theta_{ij}=\theta_{i}-\theta_{j}<\pi/2 $ if $ a_{ij}=1 $).
These solutions can be put in two categories;
i)~ Full synchronized state with $r=1$ ($ \Delta\theta_{ij}=0  $ for any $ i,j $);
ii)~ Phase-locked state with regular arrangement of phases around the phase circle with
non-zero phase difference $ \Delta\theta $, for which $r=0$. \\
The phase-locked states represent helical wave phase modulations
and their  number  depends on $ N $ and  $ k $. For instants, in
the case of $ N=1000 $  and $ k=10 $,  there are $ 10 $ of such
states with nearest neighbor  phase differences  $
\Delta\theta_{nn}^{\alpha}=2\pi/\lambda^{\alpha}$, in which    $
\lambda^{\alpha}=20,25,40,50,100,125,200,250,500,1000 $, are the
wavelengths  of the  helical states indicated by $
\alpha=1,2,\cdots 10 $, respectively.

\begin{figure}[t]
\begin{center}
\includegraphics[scale=0.35]{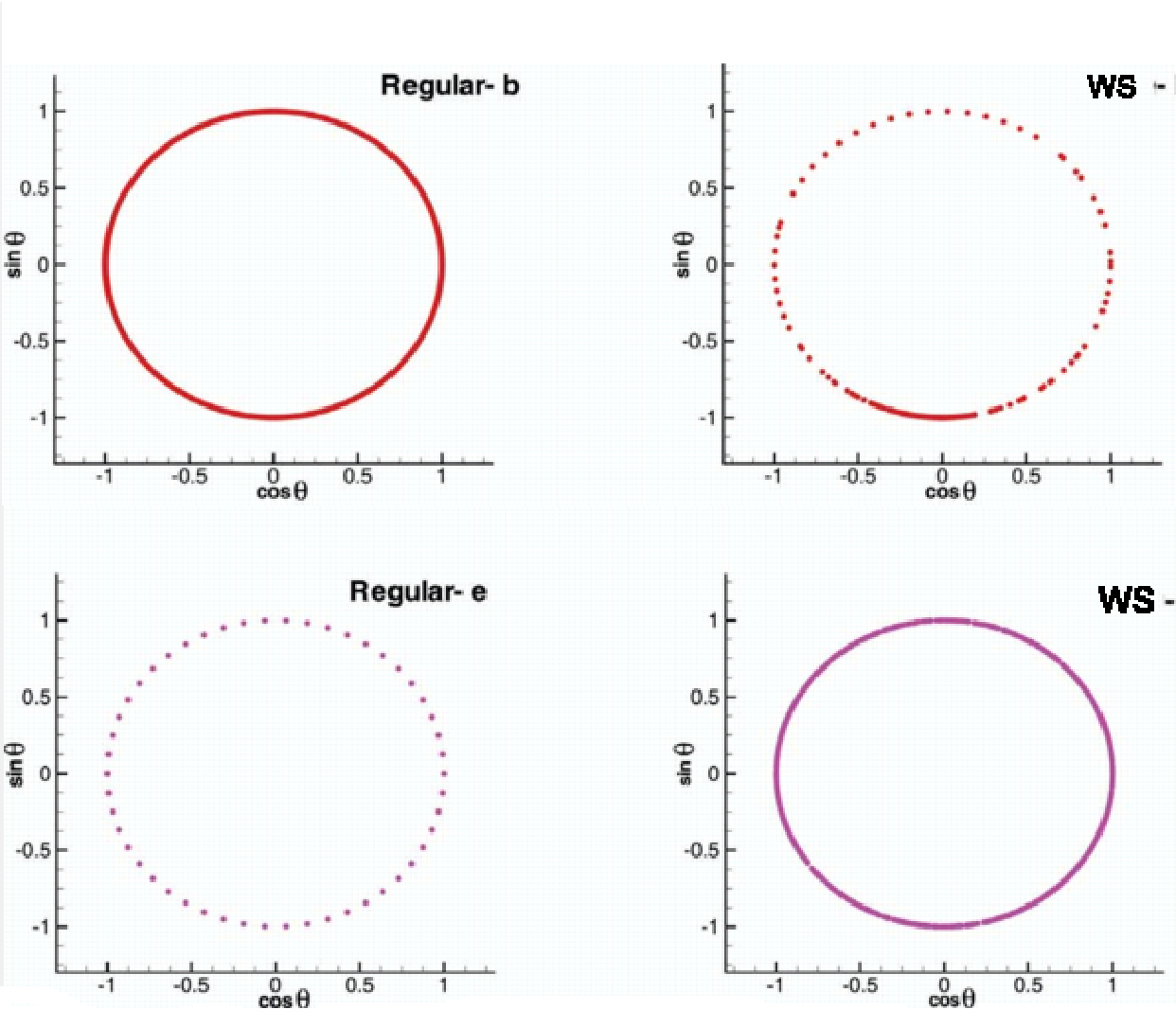}
\caption{(Color on-line) steady state phase configurations  in  two
 helical patterns of regular network and corresponding steady
states of WS networks with $N=1000$ and $\langle k \rangle=10$.
(b)$ \lambda=1000 $, and
(e)$ \lambda=50 $. $ \lambda $  is the wavelength  of helical
states in regular network. } \label{fig2}
\end{center}
\end{figure}
The stationary phase configuration  of all nodes, corresponding to
the initial conditions in Fig.(\ref{fig1}), are plotted in
Fig.(\ref{fig2}), both for the regular and  its offspring  WS
network.  This plots corresponds to the helical patterns with
phase differences $ \lambda=1000,50$, denoted in
Fig.(\ref{fig1}) by indices, b and  e, respectively. This
figure shows, rewiring a  regular network with phase-locked state,
deforms its    helical pattern    to  an inhomogeneous  state in
the subsequent WS one. Therefore, a WS network possesses various
stable  stationary states whose number equals the number of
helical patterns in its parent regular network.

\begin{figure}[t]
\begin{center}
\includegraphics[scale=0.7]{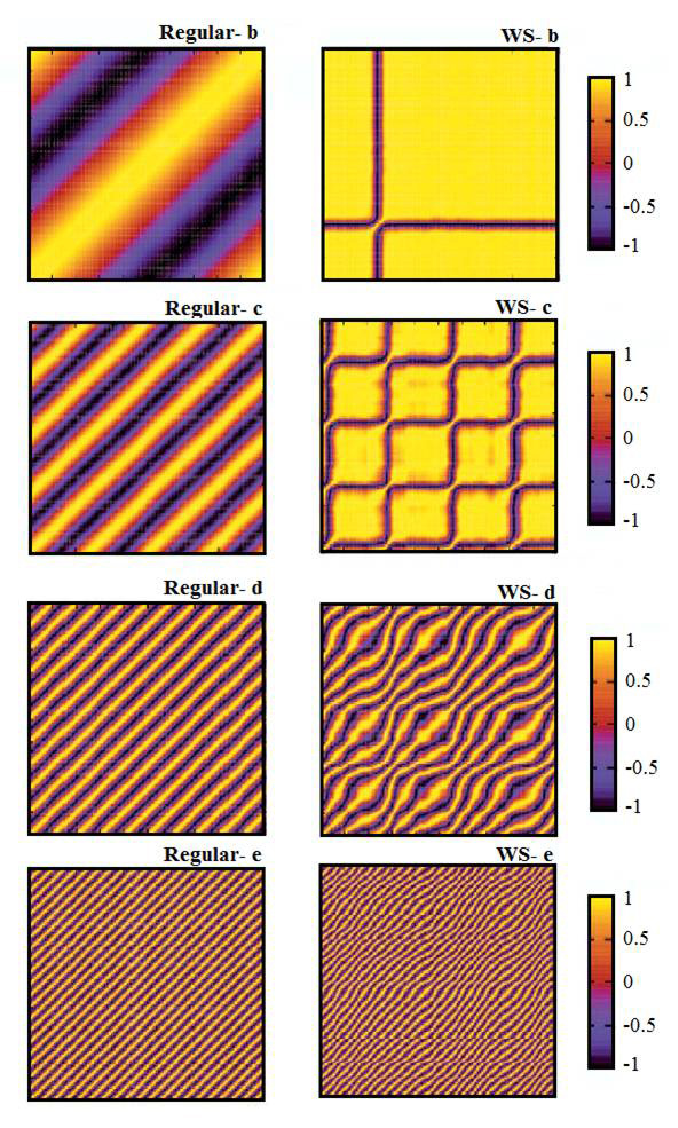}
\caption{(Color on-line) Density plot of correlation matrix
elements ($D_{ij}$) for  $ 4 $  helical states of  regular network
and corresponding stationary states in WS network  with $N=1000$
and $\langle k \rangle=10$.  (b)$ \lambda=1000 $,
(c)$ \lambda=250$, (d)$ \lambda=100 $ and (e) $ \lambda=50 $. $\lambda $  is the
wavelength  of helical states in regular network.  } \label{fig3}
\end{center}
\end{figure}
The local structure of the steady state can be better clarified by the correlation matrix $D$ defined as~\cite{D}:
\be
    D_{ij}=\lim_{\Delta t\rightarrow \infty}
    \frac{1}{\Delta t}\int_{t_r}^{t_r+\Delta t}
    \cos(\theta_i(t)-\theta_j(t))dt,
    \label{Dmatrix}
    \ee
in which $t_r$ is the time needed for  reaching  to stationary
state. The matrix element $-1 \leq D_{ij}\leq 1$ is a measure of
coherency between each pair of nodes. In the case of full
synchrony between  $i$ and $j$ ($\theta_{i}=\theta_{j} $) the
correlation matrix element is   $D_{ij}=1$  and in the  case of
anti-phase locking  ($\theta_{i}-\theta_{j}=\pi $), the value of
matrix element  is  $D_{ij}=-1$. Fig.(\ref{fig3}) represents  the
density plot of correlation matrix elements  for  the four
steady states of regular and  WS networks corresponding to
Figs.(\ref{fig1}) and (\ref{fig2}). This plots  clearly show the
inhomogeneous structure of the helical patterns before and after
rewiring of the regular network.  The correlation matrix
represents strip structures in its  density plot for helical
states in regular network and  the width of the strips are
proportional to the wavelength of the helices.  One  can also
observe   from these plots that  converting  the regular network
to WS,  the helical patterns are substantially affected,  provided
$ \lambda $ being large.  The strip structure  of matrix $ D $ is
almost preserved for small wavelengths,  indicating  the small
wavelength helical patterns, despite of little  deformations,  are
stable against rewiring of network. For large wavelength patterns
of regular network, the majority of nodes in the corresponding  WS
phase configurations  are synchronized with each other, however
there are some isolated nodes  in Anti-phase locking  with the
rest. These isolated nodes are {\it topological defects}  and
induce  spiral phase textures around them, in such a way that the
phase  of surrounding  oscillators  varies continuously from $0$
to $\pi$,  by getting away from these nodes. The number of these
defects increases by  decreasing   the wavelength of corresponding
helical pattern.  For example, it can be seen from
Fig.(\ref{fig3}) that for  $\lambda=1000 $ there is one  while for
$ \lambda=250 $ there are four point defects. Once the structure
of the steady states of deterministic Kuramoto model on WS network
is known, it would be interesting to investigate the effect of
noise on such states.

\section{The effect of random force}

\begin{figure}[t]
\begin{center}
\includegraphics[scale=0.3]{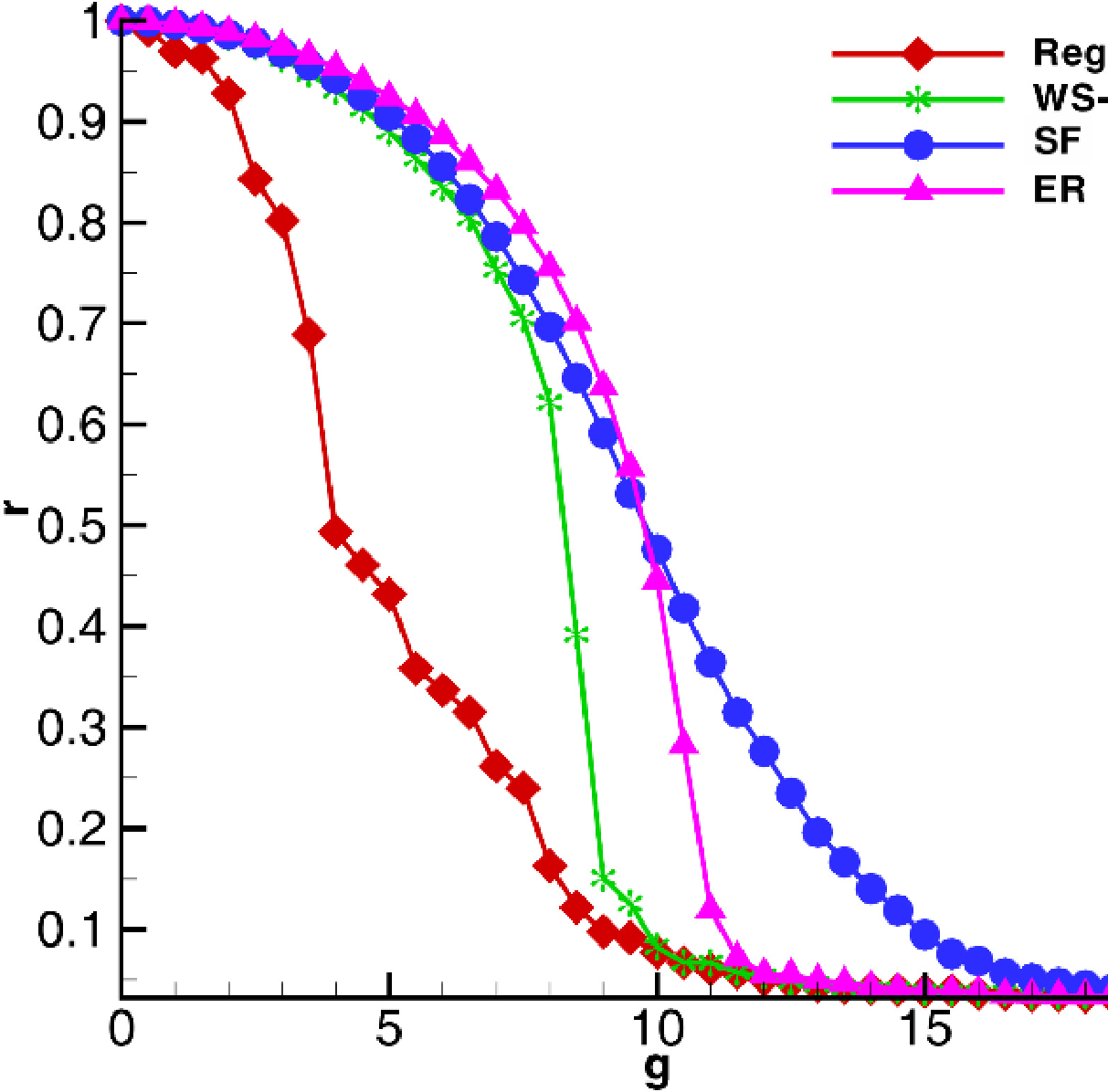}
\includegraphics[scale=0.37]{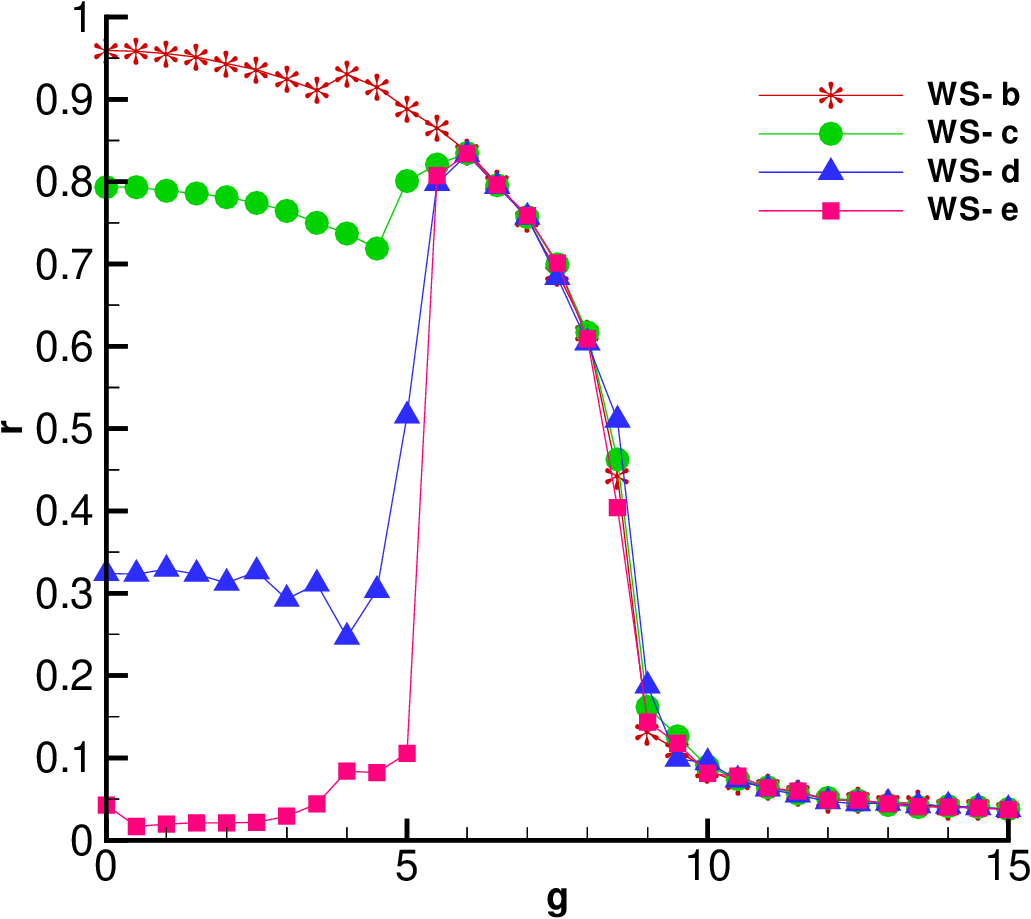}
\caption{(Color on-line) Stationary order parameter versus
reduced noise intensity for the four network types.  (top panel):
Regular, ER, SF and Full synchronized state of WS. (bottom panel):
Four phase locked states of WS corresponding to states represented
in Fig.(\ref{fig3}).  Number of nodes and mean degree  for the
three networks are $N=1000$ and $\langle k \rangle=10$,
respectively.} \label{fig4}
\end{center}
\end{figure}
A network of oscillators could be  plagued by some external random
forces. The effect of such forces may be modelled by an uncorrelated  white noise
$(\eta_i(t))$   applying  to all nodes. Adding this noise to  Eq.(\ref{kuramoto2}), we have:
    \be
    \dot{\theta}_i=K\sum_{j=1}^{N}a_{ij}
    \sin(\theta_j-\theta_i)+\eta_i(t)\; ,i=1,\ldots,N\,,
    \label{noise}
    \ee
where $ \br\eta_i(t)\kt=0 ,
\br\eta_i(t)\eta_j(t')\kt=2D\delta(t-t')\delta_{ij} $ with $D$
being the  variance or  intensity of the noise. In our numerical
work, we  choose a box distribution in the interval $[-w/2,w/2]$
for  $\eta$, so that its variance is equal to $D=w^2/24$. It can
be shown that by proper re-scaling of the  time variable, the
effect of  parameters $D$ and $K$ can be included in a single
parameter  $g^2=\frac{D}{K}$ ~\cite{khoshbakht}, converting  the
dynamical equations to:
    \be
    \frac{d\theta_i}{d\tau}=\sum_{j=1}^{N}a_{ij}
    \sin(\theta_j-\theta_i)+g\xi_i(\tau)
    \label{noise2}
    \ee
where $\tau=Kt$ is the re-scaled time variable and
$\xi_i(\tau):=\eta_i(t)/g$ is a random variable in the interval $
[-1/2,1/2 ] $.

The numerical integration of  Eq.(\ref{noise2}) is carried out by
employing  Euler method for its deterministic part and Ito's
algorithm~\cite{ito} for the  stochastic part. Fig.(\ref{fig4})
represents  the variations of stationary   order parameters
($r(\infty)$)   versus re-scaled  noise intensity  $g$, for the
four  network types regular, SF, ER and WS (top panel for fully
synchronized state and bottom panel for phase-locked ones).

\begin{figure}
\begin{center}
\includegraphics[scale=0.7]{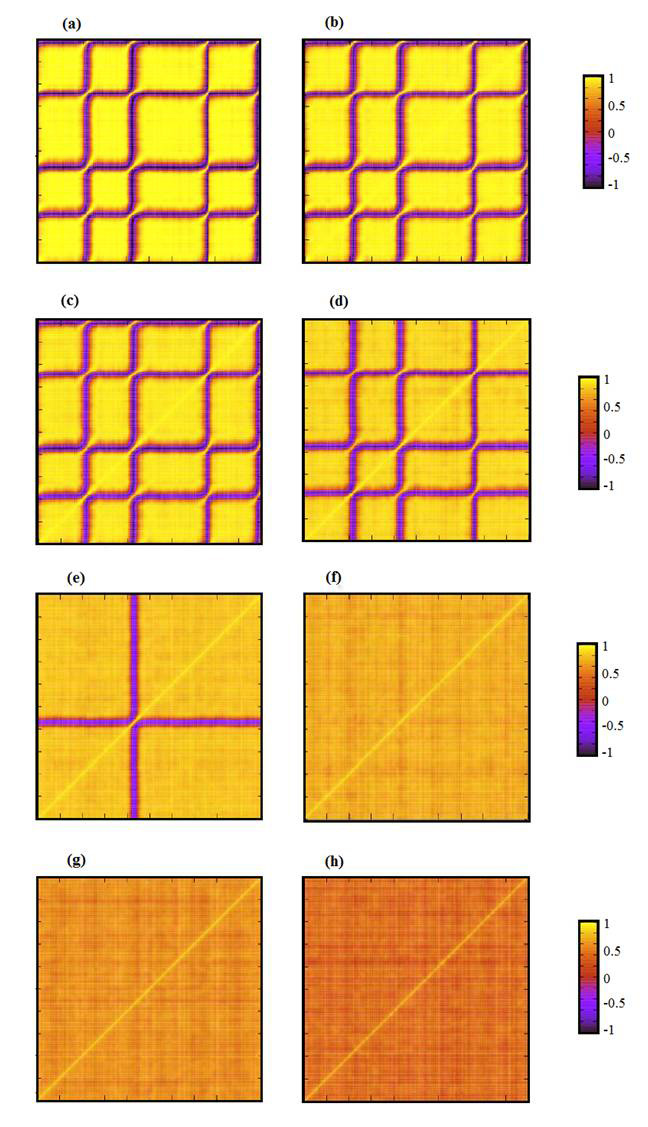}
\caption{(Color on-line) Density plot of correlation matrix
elements ($D_{ij}$) for a steady state with four point defects on
SW network and different   noise intensities, (a) $ g=0 $, (b) $
g=2 $, (c) $ g=3 $, (d) $ g=4 $, (e) $ g=5$, (f) $ g=6 $,  (g) $
g=7 $ and (h) $ g=8 $. Number of nodes and mean degree  for the
three networks are $N=1000$ and $\langle k \rangle=10$,
respectively } \label{fig5}
\end{center}
\end{figure}
\begin{figure}[t]
\begin{center}
\includegraphics[scale=0.47]{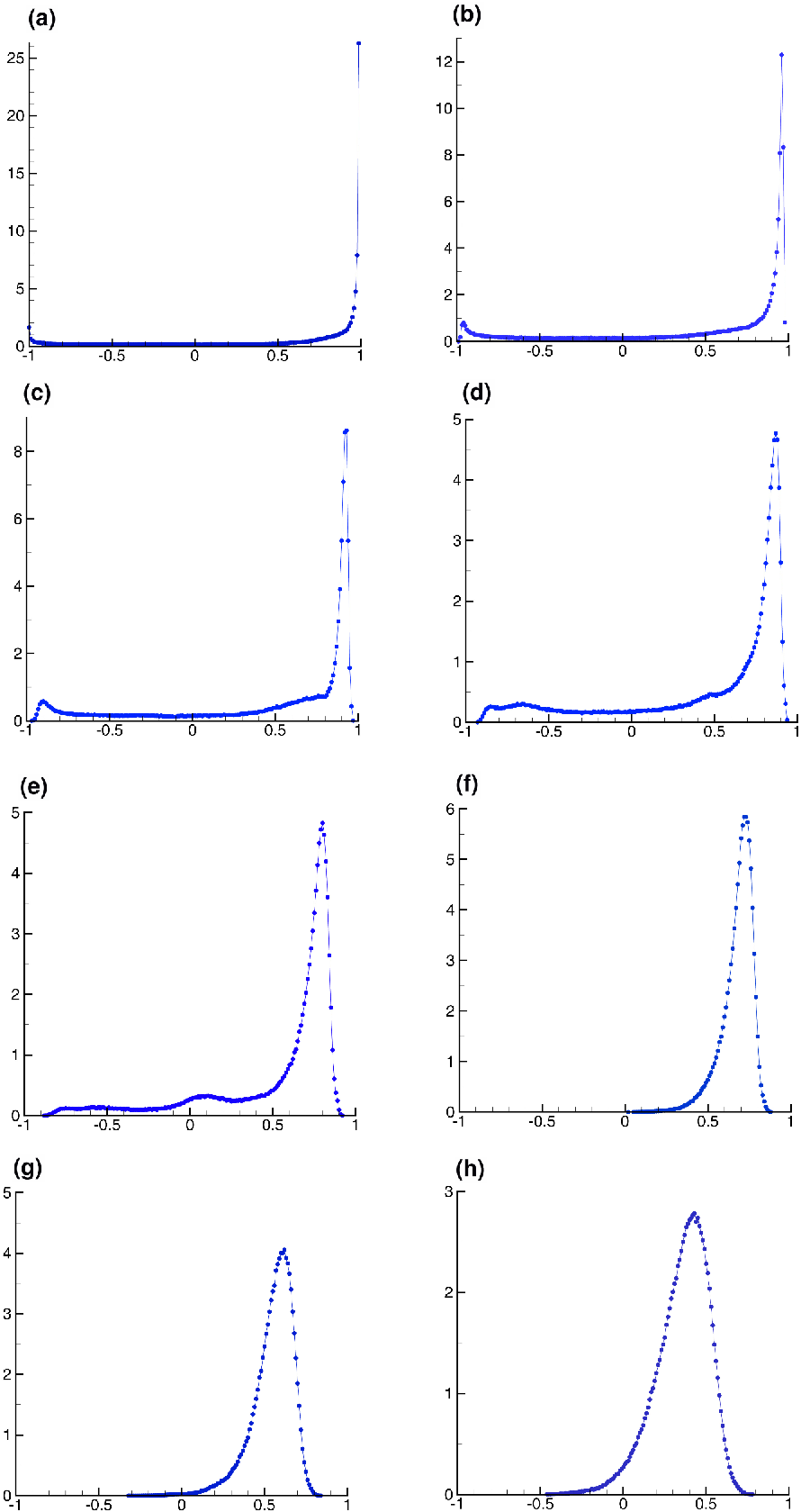}
\caption{(Color on-line) Probability distribution function of
correlation matrix elements ($D_{ij}$) for different  noise
intensities: (a) $ g=0 $, (b) $ g=2 $, (c) $ g=3 $, (d) $ g=4 $,
(e) $ g=5 $, (f) $ g=6 $, (g) $ g=7 $ and (h) $ g=8 $. Number of
nodes and mean degree  for the three networks are $N=1000$ and
$\langle k \rangle=10$, respectively.} \label{fig6}
\end{center}
\end{figure}
\begin{figure}[t]
\begin{center}
\includegraphics[scale=0.32]{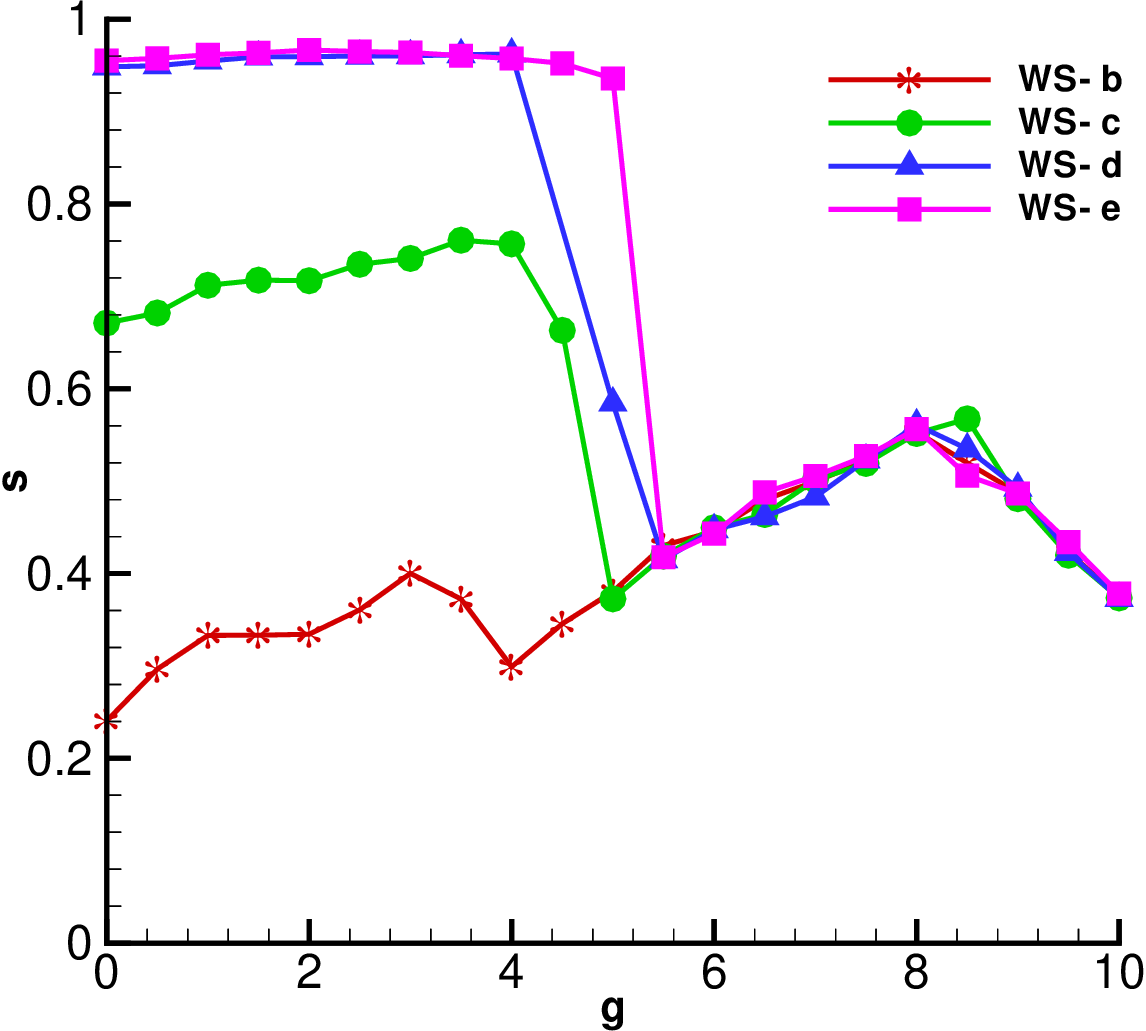}
\includegraphics[scale=0.35]{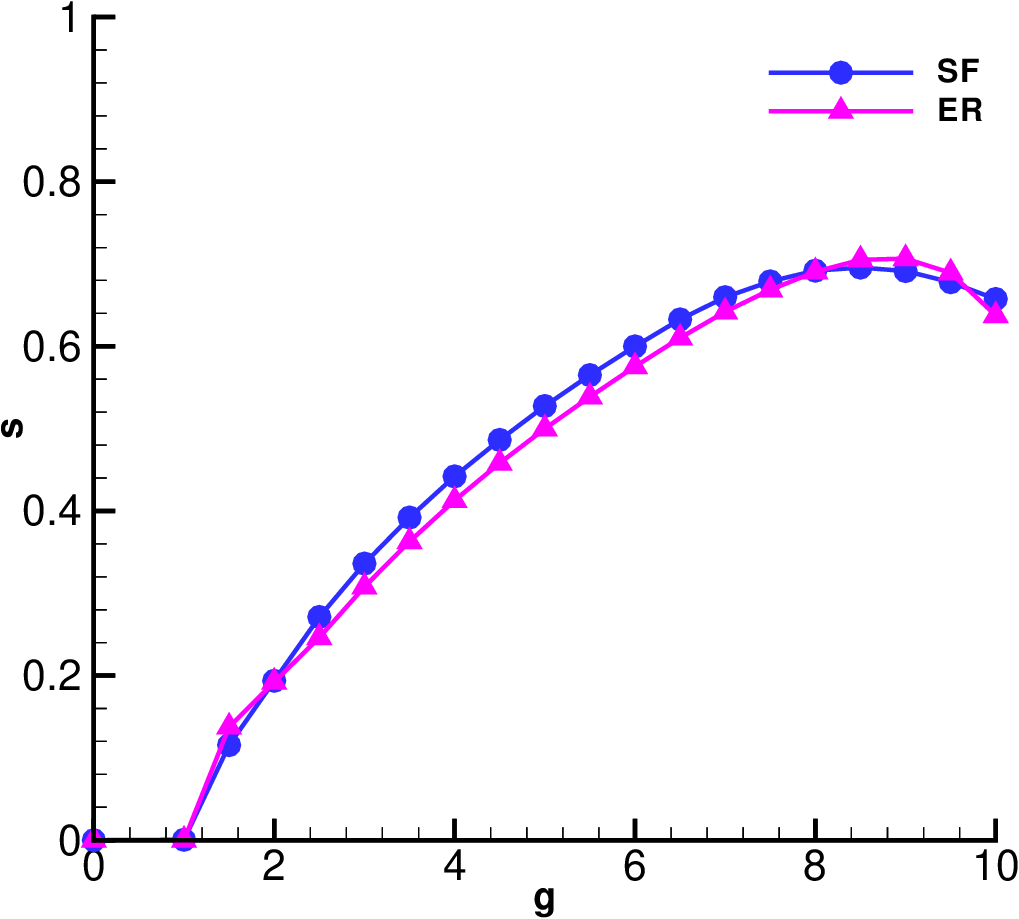}
\caption{(Color on-line) top: complexity of small world network versus
noise intensity for  $N=1000$ and $\langle k \rangle=10$. These plots are corresponding to
the initial conditions leading to the helical
state in parent regular network with wavelength $\lambda=1000, 250, 100, 10$.
bottom: complexity of ER and SF networks versus
noise intensity for  $N=1000$ and $\langle k \rangle=10$.}
\label{fig7}
\end{center}
\end{figure}

 By inspecting this figure, one can extract two essential results:
(i)  As can be seen from the top panel, for all the four networks,
the order parameter starting from full synchronized state with
$r({\infty})=1$, monotonically decreases when the noise is turned
on. The critical coupling ($g_c$) at which the synchrony
disappears among  the oscillators is the greatest for SF and the
smallest for regular network. Therefore  the coherent state in the
SF  lasts longer against noise than  ER, WS  and regular with the
same average degree and coupling constant.  The more fragility of
fully synchronized state in regular and WS networks can be
explained in terms of the formation of some local clusters  in
these networks.  The phase differences   among   different
clusters of oscillators   tend to become large  by the  effect of
random forcing, hence leading to rapid fall  of the order
parameter.  The persistence of synchronized state in SF network
has been argued to be  related to  the  existence of few nodes
with very large number of connections (hubs) in this type of
networks~\cite{khoshbakht}.

(ii) The bottom panel of Fig.(\ref{fig4}) shows that for
inhomogeneous  phase locked states in WS network, variation of
$r(\infty)$ versus noise is non-monotonic.  It remains almost
constant for small noise strengths and reaches to a maximum in  an
interval of reduced noise intensity. Therefore in  these cases,
instead of playing destruction role,  noise promotes  the
synchrony among the oscillators.

The noise-induced synchronization is also called {\it stochastic synchronization} and its occurrence
in WS networks can be explained in terms of  defect patterns  in the steady states of the Kuramoto model.
Fig.(\ref{fig5}) represents the evolution of correlation matrix density plots  versus reduced noise intensity,
$ g $, for a specific steady state of WS network with four topological defects.
It can be seen in this figure that turning the noise on, the defects resist against the noise up to
$ g\sim 4 $ and for $ g>4 $ they begin to disappear until $ g\sim 6 $  where they vanish completely.
 Disappearance of defects enhances the homogeneity  in the system and so the synchrony among the oscillators.
 This is more apparent   in probability distribution of correlation matrix elements ($ p(D) $) shown in
Fig.(\ref{fig6}). As can be seen in this figure,
$ p(D) $ has two peaks at $ D=1,-1 $ for $ g=0 $. Increasing the noise intensity the two peaks  move toward
each other and at the onset of stochastic synchronization, $ g\sim6 $,  they emerge in one peak.
At this point the variance of $ p(D) $ reaches to its minimum and again  rises  by increasing the noise strength.
 Fig.(\ref{fig7}) represents the complexity of WS, ER and SF networks in terms of reduced noise intensity.
 The complexity is defined by Shannon entropy of $p(D)$~\cite{complexity}:
\begin{equation}
S=\left(-\sum_{i=1}^{m} p_{i}\ln p_{i}\right)/\ln m,
\end{equation}
in which $m$ is the number of bins in division of $p(D)$ ($m=200$
in this work). This quantity measures both the ability of a
network to synchronize as a whole (integration) and in the mean
while preserving the independence of its subsystems (segregation).
This intermediate regime with high complexity  is desirable for
functioning of real neural networks. 

In  top panel of
Fig.(\ref{fig7}), we  compare the complexity of Kuramoto model on
WS network for different initial conditions, leading to the
patterns  shown in Fig.(\ref{fig3}). This plot shows that the
complexity  is larger for the patterns with greater spatial
inhomogeneity (corresponding to the helical patterns with smaller
wavelength  in the regular networks), and these patterns are more
robust against noise. By applying the noise, the complexity
remains more or less unchanged until the the onset of stochastic
synchronization at which shows a sudden fall. Beyond this point,
the complexity tends to rise and reaches to a maximum at the noise
strength by which the synchronization vanishes. On the  contrary,
for ER and SF networks, shown in the bottom panel of
Fig.(\ref{fig7}), the complexity monotonically raises with noise
strength and reaches to a maximum at the onset of  the vanishing
of synchronization. 

\begin{figure}[b]
\begin{center}
\includegraphics[scale=0.4]{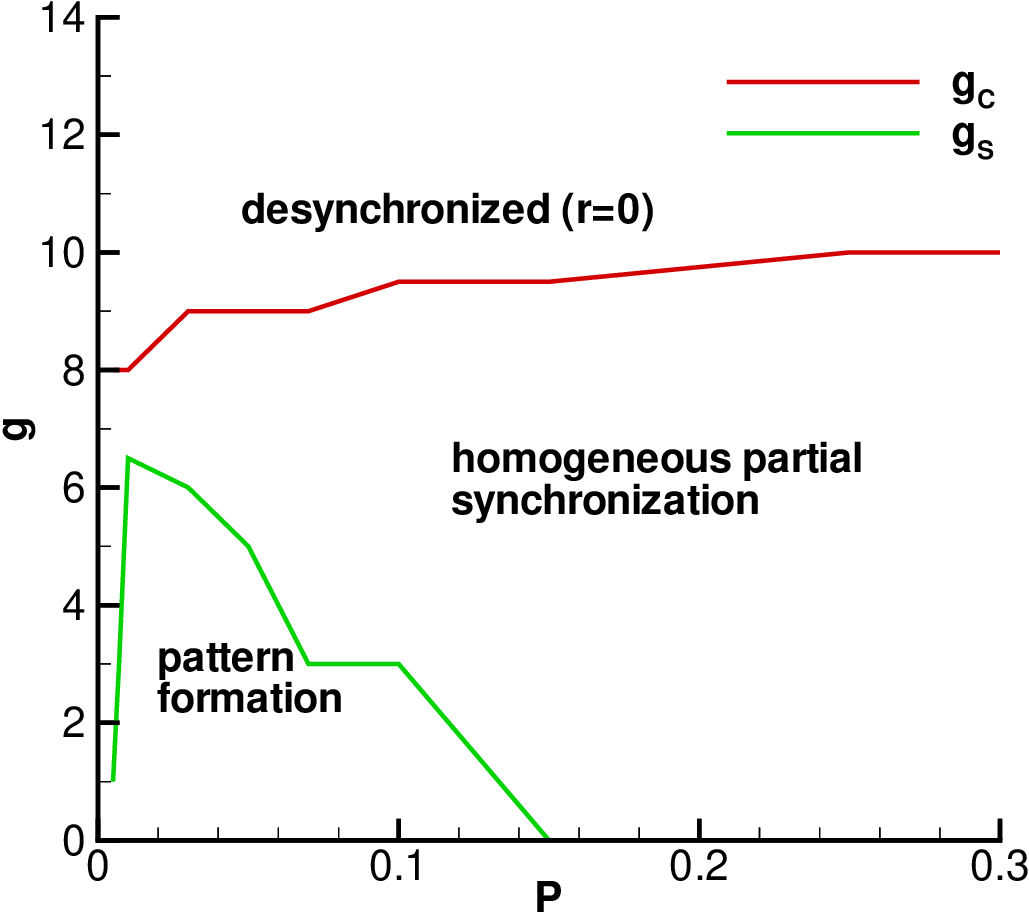}
\caption{(Color on-line) Phase diagram of noisy Kuramoto model on a SW network
with  $N=1000$ and $\langle k \rangle=10$, in $g-p$ space.
$ g_{s} $ (light-gray line) and  $ g_{c} $ (dark-gray line) denote the noise strengths
at the onsets of stochastic synchronization  and desynchronization,
respectively.}
 \label{fig8}
\end{center}
\end{figure}

Finally,  we investigate  the occurrence of stochastic
synchronization in terms of link rewiring probability $p$. We
found that  value of the noise strength at which the stochastic
synchronization occurs, reaches to maximum at $p\sim 0.02$ and
then decreases with increasing $p$ and vanishes  at $p\sim 0.17$.
For $p>0.17$, the behavior of the noisy  Kuramoto dynamics on SW
network is similar to  the  random networks. In Fig.(\ref{fig8})
we depict  the  phase diagram for a  SW network with $N=1000$ and
$\langle k \rangle=10$,  in $g-p$ space.

\section{conclusion}
In summary, we found  that a SW network  of similar phase
oscillators communicating with each other by Kuramoto coupling
shows novel behaviors. Unlike ER and SF networks, this system
fails to reach a full synchronized state  for any arbitrary initial
conditions. Moreover, driving it by an uncorrelated white noise,
reveals the occurrence of stochastic synchronization, a phenomenon
through which a random force  induces synchrony among the
oscillators. We discussed that the reason for this phenomenon is
laid in the stable helical patterns  in the regular networks from
which the SW ones is built. Rewiring of a regular network of
similar phase oscillators  with periodic helical pattern, ends to
complex inhomogeneous  states in the resulting SW network. The
existence of such stable inhomogeneous patterns in SW network,
appearing some times as topological point defects and also as
aperiodic helical patterns, prevents the network from reaching to
full synchrony. These patterns persist against  the noise for
small noise intensities. However  the external random forces with
moderate strengths  are able to destroy  these patterns  in favor
of more homogeneous states, hence enhance the synchronization
among oscillators. We  computed  the complexity on the SW network
in the case of  the  inhomogeneous pattern formation,  and
showed that the complexity of such states are larger than the  ER
and SF networks,  for the noise strength less than the onset of
stochastic synchronization. Therefore, as a model for neural
networks, this finding shows  that  the  functioning of such
systems  can be more efficient in the presence of  a  moderate
noise. Generalization of the above results to the more realistic case in which   the
coupling constants (coefficients of periodic couplings) are normalized to 
the degree of the nodes, is currently under investigation.   We hope  our results   may shed light on the fact that why SW networks are so ubiquitous  in natural systems.

\acknowledgments
We would like to thanks S. Strogatz for enthusiastic discussions and useful comments.


\begin{thebibliography}{99}

\bibitem{order-noise}  F. Sagu{\'e}s,  J. M. Sancho, and  J. Garc{\'i}a-Ojalvo, Rev. Mod. Phys {\bf 79}, 829 (2007).

\bibitem{SR}  L. Gammaitoni,  P. H$\mathrm{\ddot{a}}$nggi,  P. Jung, and  F. Marchesoni, Rev. Mod. Phys {\bf 70}, 223 (1998).

\bibitem{CR}  B. Lindner,  J. Garc{\'i}a-Ojalvo,  A. Neiman, and  L. Schimansky-Geier, Phys. Rep. {\bf 392}, 321 (2004).

\bibitem{transport}   P. Reimann, Phys. Rep. {\bf 361}, 57  (2002).

\bibitem{transition}  W. Horsthemke, and  R. Lefever, {\it Noise-Induced Transitions}, (Springer, Berlin, 1984).

\bibitem{collective-firing} S. K{\'a}d{\'a}r,  J. Wang, and  K. Showalter, Nature (London) {\bf 391}, 770 (1998);
 S. Alonso,  I. Sendi$\mathrm{\tilde{n}}$a-Nadal,  V. P$\mathrm{\acute{e}}$rez-Mu$\mathrm{\tilde{n}}$uzuri,  J. M. Sancho, and
F. Sagu$\mathrm{\acute{e}}$s , Phys. Rev. Lett. {\bf 87} 078302 (2001);
S. Tanabe, and  K. Pakdaman, Biol. Cybern. {\bf 85} 269 (2001);
 C. J.  Tessone,  A. Scir$\mathrm{\grave{e}}$ ,  R. Toral , and  P. Colet, Phys.  Rev.  E {\bf 75} 016203 (2007).

\bibitem{noise-neuron}  W. H. Calvin, and  C. F. Stevens,  Science {\bf 155} 842 (1967).

\bibitem{brain-sync}
G. B. Ermentrout,  R. F. Gal{\'a}n, and  N. N. Urban, Trends in Neuroscience {\bf 31} 428 (2008).

\bibitem{signal-neuron}
 A. N. Burkitt, and  G. M. Clark, Neural Comput. {\bf 11}, 871 (1999);
 E. Salinas, and  T. J. Sejnowski,  Nat. Rev. Neurosci. {\bf 2}, 539 (2001);
 A. D. Reyes, Nat. Neurosci. {\bf 6}, 593 (2003);
 P. H. Tiesinga, and  T. J. Sejnowski, Neural Comput. {\bf 16}, 251 (2004).

\bibitem{code-sync}
 M. Stopfer {\it et al}, Nature {\bf 390}, 70 (1997);
A. K. Engel {\it et al}, Conscious Cogn. {\bf 8}, 128 (1999);
 M. N. Shadlen, and  J. A. Movshon, Neuron {\bf 24}, 67 (1999);
J. A. Movshon, Neuron {\bf 27}, 412 (2000).

\bibitem{noise-gene}
 M. Springer, and  J. Paulsson, Nature {\bf 439}, 27 (2006);
 T. Zhou,  L. Chen,  and  K. Aihara,  Phys. Rev. Lett. {\bf 95}, 178103 (2005).

\bibitem{rev-network}   A. Arenas, A. Diaz-Guilera, J. Kurths, Y. Moreno and C. Zhou, Physics Reports  {\bf 469}, 93-153 (2008).

\bibitem{kuramoto}  Y. Kuramoto, Lecture Notes Physics (Springer, New York,
1975), Vol. 39, pp. 420-422;
 Y. Kuramoto, {\it Chemical Oscillations, Waves, and Turbulence}  (Springer, Berlin, 1984).

\bibitem{skuramoto}
 J. A. Acebr\'{o}n,  L. L. Bonilla, and
Conrad J. P\'{e}rez Vicente,  F\'{e}lix Ritort, and  R. spigler, Rev.~Mod.~Phys. {\bf 77}, 137 (2005);
 B. C. Bag,  K. G. Petrosyan, and Hu Chin-Kun, Phys.  Rev. E {\bf 76}, 056210 (2007);

\bibitem{khoshbakht}  H. Khoshbakht,  F. Shahbazi, and,  K. Aghababaei Samani, J. Stat. Mech,  P10020 (2008).


\bibitem{watts} J. D.  Watts, and  S. H. Strogatz, Nature {\bf 393}, 440  (1998).

\bibitem{strogatz}  S. H. Strogatz, Nature {\bf 410}, 268 (2001).



\bibitem{sw-sr}
 Z. Gao,  B. Hu, and  G. Hu, Phys. Rev. E {\bf 65}, 016209 (2001);
 H. Hong,  B. J. Kim, and  M. Y. Choi, {\it ibid}. {\bf 66}, 011107  (2002) ;
 M. Perc, and  M. Gosak, New J. Phys. {\bf 10}, 053008  (2008) .

\bibitem{sw-cr}
 O. Kwon, and  H.-T.  Moon, Phys. Lett A {\bf 298}, 319 (2002);
 O. Kwon,  H.-H. Jo , and  H.-T. Moon , Phys. Rev. E  {\bf 72}, 066121(2005).

\bibitem{sw-brain}
D. S. Bassett, and  E. Bullmore, The Neuroscientist {\bf 12}, 512 (2006);
E. Bullmore, and  O. Sporns, Nature Reviews Neuroscience {\bf 10}, 186 (2009).


\bibitem{sw-synch}  H. Hong, M. Y. Choi, and  B. J. Kim,  Phys. Rev. E {\bf 65}, 026139 (2002).

\bibitem{BA}  A. L. Barab\'{a}si, and  R. Albert, Science {\bf 286}, 509 (1999);
A. L.  Barabasi,  R. Albert, and  H. Joeng,  Physica A {\bf 272}, 173 (1999).

\bibitem{ER}  P. Erd\"{o}s,  and  A. R\'{e}nyi, Publ. Math. Debrecen {\bf 6}, 290  (1959);
 Publ. Math. Inst. Hung. Acad. Sci {\bf 5}, 17 (1960).

\bibitem{reg-basin} A. D. Wiley, S. H. Strogatz, and M. Girvan, Chaos {\bf 16}, 015103 (2006) .

\bibitem{D}  J. G{\'o}mez-Garde$\mathrm{\tilde{n}}$es,  Y. Moreno, and A. Arenas, Phys. Rev. Lett. {\bf 98},  034101 (2007).

\bibitem{ito}  C. W. Gardiner, {\it Handbook of Stochastic
Methods for Physics, Chemistry and the Natural Sciences},
(Springer-Verlag, 1980).

\bibitem{complexity} M. Zhao, C. Zhou, Y. Chen, B. Hu, and B-H. Wang, Phys. Rev E {\bf 82}, 046225 (2010).



















\end{thebibliography}
\end{document}